\documentclass[letterpaper,10pt,final,balancelastpage,floats,aps,twocolumn]{revtex4}
\usepackage{amsmath}
\usepackage{amssymb}
\usepackage{multirow}
\usepackage{graphicx}

\begin{document}

\title{Targeted Excited State Algorithms}
\author{Jonathan J. Dorando, Johannes Hachmann, and Garnet Kin-Lic Chan}
\affiliation{Department of Chemistry and Chemical Biology, Cornell
University, Ithaca, NY 14853-1301, USA}

\begin{abstract}

To overcome the limitations of the traditional state-averaging
approaches  in excited state
calculations, where one solves for and represents all states
between the ground state and excited
state of interest, we have investigated a number of new
excited state algorithms. Building on the work of van der Vorst and
Sleijpen (SIAM J. Matrix Anal. Appl., \textbf{17}, 401 (1996)), we
have
implemented   Harmonic Davidson and State-Averaged Harmonic Davidson algorithms
 within the context of the Density Matrix Renormalization
Group (DMRG).  We have assessed their accuracy and stability of
convergence in  complete  active space DMRG calculations on the
low-lying excited states in the acenes ranging from naphthalene to
pentacene.  We find that both algorithms offer increased accuracy
over the traditional State-Averaged Davidson approach, and in
particular, the State-Averaged Harmonic Davidson algorithm offers an
optimal combination of accuracy and stability in convergence.


\end{abstract}

\maketitle

\section{Introduction}

Many excited states possess complicated electronic structure which
cannot be described by a single dominant electronic
configuration.  For such states, a reliable
description requires a multireference quantum
chemistry method.

Recently, the Density Matrix Renormalization Group (DMRG) has
emerged as a new tool for multireference quantum chemistry problems
\cite{yaron1998cod, white1998aiq, legeza2003cad,
fano1998dmr,shuai1997qce, chan2002hcc, hallberg2006rev}. When
applied to bond-breaking, it achieves a balanced description across
potential energy curves due to its reference-free nature
\cite{legeza2003qds, moritz2005rdc, chan2004sad}. Reduced-scaling
DMRG algorithms have also been developed and applied to large
multireference problems in quasi-one-dimensional systems such as
conjugated polyenes and acenes \cite{hachmann2006mcl,
hachmann2007dirad}.

The DMRG ansatz can be written as a linear expansion in terms of    many-body
functions which are subsequently optimised with respect to  internal
non-linear degrees of freedom $\{ \mathbf{R} \}$,
\begin{equation}
|\Psi\rangle  = \sum_{lr} \psi_{lr} \left |lr  \left(\{ \mathbf{R} \} \right)\right \rangle \label{eq:basicansatz}
\end{equation}
Note that if we choose the expansion functions $|lr\rangle $ to be Slater determinants  and the internal degrees of freedom $\{
 \mathbf{R} \}$  to be their constituent orbitals,  the above ansatz
 describes the Complete-Active-Space Self-Consistent-Field (CASSCF) wavefunction \cite{casscfrev}.
In the DMRG, the expansion functions are instead complicated
many-body basis states and the non-linear degrees
 of freedom are renormalisation matrices, which allows for
a particularly compact and efficient expansion
\cite{schollwock2005dmr}.

To obtain excited  states in the DMRG we usually use the iterative
Davidson algorithm to solve for
 eigenvectors $|\Psi_i \rangle = \psi_{lr}^i|lr\rangle$ ranging  from the ground-state  to the excited state
of interest \cite{hallberg2003dmr}. The non-linear parameters $\{
\mathbf{R} \}$ for  these states are subsequently optimised for  a
 density matrix that is averaged over all the  states $|\Psi_i\rangle$.
State-averaging is necessary to improve the stability of the
non-linear optimisation  and to prevent   root-flipping, which
occurs when the approximate wavefunction leaves  the convergence
basin of the target excited state and enters that of a different
excited state \cite{docken1972lpc, khait1995ssp, knowles1992,
  hoffmann2002, cances2006}.

The drawbacks of this conventional approach, which we shall refer to as the
State-Averaged Davidson (SA-D) algorithm,  become
clear if one is interested in higher regions of the spectrum because
it becomes  infeasible, both in terms of computational cost and
accuracy, to solve for and adequately represent
all the lower-lying eigenvectors in the state-averaged DMRG basis.
 Consequently, it is desirable to explore alternative
 algorithms that  directly yield individual or a few excited
state  wavefunctions at a time. Any such an algorithm
should also retain the stability of the SA-D algorithm during non-linear optimisation, so as to be
able to rapidly converge to the desired target excited state(s)
without root-flipping.

Iterative methods for linear algebra that work with shifted and
inverted operators such as $(\omega - H)^{-1}$  have long been used
in numerical analysis to obtain the interior (i.e. excited state)
eigenvalues of matrices \cite{morgan1991,bai2000tsa}.  Sleijpen and
van der Vorst proposed an efficient modification that used a shifted
and inverted operator to directly calculate \textit{harmonic Ritz}
approximations to excited eigenvalues and eigenvectors
\cite{sleijpen1996jdi}. We shall refer to this variant as the
\textit{Harmonic Davidson} (HD) algorithm to distinguish it from the
original algorithm above. Aside from a demonstration for the
one-electron Kohn-Sham equation in Ref. \cite{tackett2002}, we are
not aware of the application of this technique elsewhere in quantum
chemistry.

The purpose of this work is to investigate  the Harmonic Davidson
algorithm  as a means to directly  target individual excited states
and regions of the spectrum within the DMRG. One area in which the
current application to quantum chemistry differs from previous
numerical applications is the presence of a subsequent nonlinear
optimisation step for  the wavefunction. We investigate how
combining  the Harmonic Davidson procedure with state-averaging over
nearby states in the  spectrum (State-Averaged Harmonic Davidson, or
SA-HD) can be used to confer stability in this non-linear
optimisation. While we have focused on the  DMRG method here, our
findings are relevant to excited state algorithms for  other quantum
chemistry methods whose ansatz contains both linear and non-linear
parameters,  such as in the CASSCF method.

The structure of this paper is as follows. In Sec. \ref{sec:the}, we
briefly review the DMRG method and the Davidson and  Harmonic Davidson
algorithms.  In Sec. \ref{sec:app}, we present  DMRG
 calculations  on
the excited states of acenes from naphthalene to pentacene
using both direct targeting with the Harmonic Davidson
 algorithm (in both state-averaged and non-state-averaged forms)  as
 well as with the traditional (state-averaged) Davidson
 approach. We also compare our excited state spectrum with that obtained from
Equation of Motion Coupled Cluster theory. We summarise our findings in Sec. \ref{sec:con}.

\section{Theory}
\label{sec:the}
\subsection{DMRG}
\label{sec:DDA}

The quantum chemistry DMRG algorithm used in this work has been
described fully elsewhere \cite{hachmann2006mcl, chan2004als}. As a
detailed understanding is not necessary here, we shall restrict
ourselves to only the essentials.
 As described above, the DMRG wavefunction may be
 written in the form (\ref{eq:basicansatz}). The DMRG sweep algorithm
then provides an iterative method through which the many-body basis
 functions $|l\rangle, |r\rangle$ may be optimised with respect  to
 a set of  internal non-linear parameters $\mathbf{R}$. For each orbital in
 the problem we can associate an $\mathbf{R}$
 matrix, which  describes a many-body renormalisation transformation
 involving the orbital (i.e. not simply
  an orbital rotation).
In a  sweep to optimize the $|l\rangle$ states (an analogous
procedure holds for the $|r\rangle$ states),   $\mathbf{R}$
matrices are determined  from the $M$ eigenvectors of the many-particle
reduced density matrix with the largest eigenvalues. In the
ground-state case, the density matrix that determines the $|l\rangle$ states is obtained by tracing
out the $|r\rangle$ states from the
wavefunction, viz
\begin{align}
\Gamma_{ll^\prime} &= \sum_{r} \psi_{lr} \psi_{l^\prime r}
\label{eq:gamm1} \\
\Gamma_{ll^\prime} R_{l^\prime m} &= \gamma_l R_{l m}, \ \ m=1, \ldots, M
\end{align}
$M$ is referred to as the size of the DMRG many-body basis, and as $M$
increases, the DMRG wavefunction becomes exact. For excited state
calculations, it is usual to employ state-averaging to increase the
stability of the non-linear optimisation. This consists of  using an averaged reduced density
matrix in eq. (\ref{eq:gamm1})
\begin{equation}
\Gamma_{ll^\prime} = \sum_{r} w_i \psi_{lr}^i \psi_{l^\prime r}^i \label{eq:stateaverage}
\end{equation}
where typically we choose equal weights for all the states of interest.

\subsection{The Davidson Algorithm}

The Davidson algorithm provides an efficient iterative solver for
the large number of linear coefficients in the
 expansion of the ground-state DMRG wavefunction (\ref{eq:basicansatz}) \cite{crouzeix1994, saad1992}.
 $|\Psi\rangle$ is expressed in an auxiliary  basis  $\{ \eta_i \}$
 (generated by the Davidson iterations)
\begin{align}
|\Psi \rangle &= \sum_i c_i |\eta_i \rangle \\
|\eta_i\rangle &= \eta^i_{lr} |lr\rangle
\end{align}
The coefficients $c_i$ are  determined by left-projection  with
 $\langle \eta_j|$
\begin{equation}
\sum_i  \langle \eta_j | H - E | \eta_i \rangle c_i = 0 \label{eq:davideig}
\end{equation}
where  $E$ is the approximate expectation value $\langle \psi | H |\psi\rangle / \langle \psi|\psi
\rangle$. Each  iteration of the Davidson algorithm, generates a new basis
function $|\eta\rangle$  from the current trial solution
$|\psi\rangle$ via
\begin{equation}
|\eta\rangle = (\mathrm{diag}(H) - E)^{-1} (H - E) |\psi\rangle
\end{equation}
which is then orthogonalised against and added to the subspace $\{ \eta_i
\}$.

To obtain excited state eigenvectors,  the simple generalization
known as the block Davidson or Davidson-Liu algorithm
\cite{davidson1975, olsen1990pob} is typically used. Here  a
residual vector is generated for each of the states from the
ground-state up to the target excited state. Solution of the
subspace eigenvalue equation (\ref{eq:davideig}) then yields
successive approximations to all eigenstates up to the excited state
of interest. In the subsequent non-linear optimisation of the
excited state in the DMRG algorithm, the eigenvectors obtained from
the block Davidson algorithm (i.e. from the ground-state to the
target eigenvector of interest) are all averaged together in the
density matrix (\ref{eq:stateaverage}). We shall refer to this
combined procedure as the State-Averaged Davidson, or SA-D
algorithm.

From the above, we see that the primary drawbacks of the traditional
SA-D approach are (i) computational cost - we must solve for all the
states between the ground-state and excited state of interest, and (ii) decreased
accuracy - since a single set of non-linear parameters must now represent
multiple states rather than a single state.


\subsection{The Harmonic Davidson algorithm}
\label{sec:HDA}

To avoid the need to solve for the states below the excited state of
interest as in the Davidson algorithm above, classic shift and
invert methods  map the target excited state  of the Hamiltonian $H$
onto the ground-state of a shifted and inverted operator $\Omega$
\begin{equation}\label{harmonic}
\Omega = H_\omega^{-1} = (\omega - H)^{-1}
\end{equation}


The Harmonic Davidson algorithm introduced by  Sleijpen and  van der
Vorst \cite{sleijpen1996jdi} (see also Ref. \cite{bai2000tsa} for a
clear review) extends the Davidson algorithm to work with the
operator $\Omega$ without the need to explicitly compute the
operator inverse in eqn. (\ref{harmonic}). Each iteration generates
a basis $\{ \eta_i \}$, but now we expand the target excited state
$|\Psi\rangle$ in $\{ H_w \eta_i \}$
\begin{equation}
|\Psi \rangle = \sum_i c_i |H_\omega\eta_i \rangle
\end{equation}
Left projection with $\langle  \eta_i H_\omega|$ yields a
generalized eigenvalue problem
\begin{align}
\langle \eta_j  H_\omega| (H_\omega^{-1} - E_\omega^{-1}) |H_\omega \eta_i \rangle
c_i = 0 \nonumber \\
\Rightarrow \sum_i [\langle \eta_j | H_w | \eta_i \rangle_i -
  E_\omega^{-1} \langle \eta_j
  H_\omega | H_\omega\eta_i \rangle ] c_i = 0 \label{eq:generaleig}
\end{align}
where $E_\omega^{-1}$ is the current approximation to $(\omega -
E)^{-1}$.  $E_\omega$ is known as a harmonic Ritz approximation to
the corresponding eigenvalue of $H_\omega$. From
(\ref{eq:generaleig}), we see that solving the eigenvalue equation
for
  $H_\omega^{-1}$ in the subspace $\{ H_\omega \eta_i
\}$ is equivalent to solving the eigenvalue equation for the
  non-inverted operator $H_\omega$ where the trial solution is expanded in the basis
$\{|\eta_i\rangle \}$, and the coefficients are obtained by right
projection using a \textit{different} space $\{ \langle \eta_j
  H_\omega| \}$.
This suggests that subspace $\{ \eta_i \}$ for eqn. (\ref{eq:generaleig})
can also be    generated from the trial solution $|\psi\rangle$ through a Davidson-type iteration
\begin{equation}
|\eta\rangle =  (\text{diag}({H_\omega}) - E_\omega^\prime)^{-1} (H_\omega - E_\omega^\prime) |\psi\rangle
\end{equation}
where here $E_\omega^\prime$ refers to the expectation value
$\langle \psi | H_\omega \psi \rangle / \langle \psi | \psi
\rangle$, which is distinct from $E_\omega$ appearing in eqn. (\ref{eq:generaleig}).

While we could  obtain the excited state eigenvalues and
eigenvectors directly from the generalized eigenvalue problem
(\ref{eq:generaleig}), in practice it is numerically more stable to
consider a slightly different form. By Schmidt orthogonalization, we
can construct an orthogonal decomposition $\{ \tilde{\eta}_i\}$ of
$\{ H_\omega \eta_i \}$
 such that $\langle \tilde{\eta}_j H_\omega |
H_\omega\tilde{\eta}_i \rangle = \delta_{ji}$. Re-expressing the
eigenvalue problem in this basis gives
\begin{equation}
\sum_i (\langle \tilde{\eta}_j | H_\omega | \tilde{\eta}_i \rangle-
E_\omega^{-1} \delta_{ji})
c_i = 0 \label{eq:orthoharmonic}
\end{equation}

From eqn. (\ref{eq:orthoharmonic}) we see that implementing the
Harmonic Davidson algorithm requires only  minor alterations to the
traditional Davidson routine relating to the change in the subspace
from $\{ \eta_i \}$ to $\{ \tilde{\eta}_i \}$. In essence, there are
only two additional steps: the subspace  functions  are first
multiplied by $H_\omega$, and second, they are Schmidt
orthogonalized to yield $\{ \tilde{\eta}_i \}$.

In our later DMRG calculations, we will refer to the use of the
above iterative procedure to solve for the linear coefficients
together with the non-linear optimisation of the many-body basis
functions $|l\rangle, |r\rangle$ without state-averaging,
collectively, as the  Harmonic Davidson algorithm (HD).

While the operator $H_\omega$ has the target excited state of
interest as its ground-state eigenvector, stable convergence  is not guaranteed in the
non-linear optimisation. However, the  formulation of the excited state problem as a
ground-state minimization, albeit with a different operator
$\Omega$,  illustrates that root-flipping is really no
different from the poor convergence that may be found in
difficult ground-state DMRG calculations.
Consequently, the same procedures may be used to eliminate the convergence difficulty:
either we can increase the size $M$ of the DMRG basis or we can employ
a state-average over the competing states. While we do not know
\textit{a priori} which states will cause convergence difficulties, it
is reasonable to assume that they must lie energetically near  our
state of interest. We have thus implemented two types of
State-Averaged Harmonic Davidson (SA-HD) algorithms. In the first
(referred to as simply SA-HD) we average over the first $n$ excited
states of $\Omega$. These correspond to
the $n$ excited states that lie immediately above our target excited
state in the spectrum of $H$. In the second, we average over the
$n$ states which lie closest (on either side) to the target excited
state in the $H$ spectrum. We refer to this variant algorithm as SA-HDa.

The second variant (SA-HDa) is particularly suited to an alternative
way of using the shift $\omega$. Rather than choosing a shift to
target a specific excited state, we can instead choose to find the
excited states around a given shift. If stable convergence is not
achieved, we simply then increase the number of states used in the
SA-HDa average until convergence is recovered. In this way, we can
patch together the spectrum piece by piece by using successively
higher shifts.

\section{Application to Acenes}
\label{sec:app}

We have investigated the low-lying states of the acene series
ranging from naphthalene (2-acene) to pentacene (5-acene). In the
following subsections, we describe the details of the computations
(Sec. \ref{sub:cd}), examine the  excitation energies using the
State-Averaged, Harmonic Davidson, and State-Averaged Harmonic
Davidson DMRG algorithms (Sec. \ref{sub:SADMRG}), and finally  use
the (near-exact) DMRG results to assess the accuracy of the
excitation spectrum obtained from Equation-of-Motion Coupled Cluster
theory (EOM-CC) (Sec. \ref{sub:CCDMRG}).

\subsection{Computational Details}
\label{sub:cd}

\begin{table*}[tbp]
\caption{RHF, CCSD, and DMRG(500) total energies of the acenes. All energies are in hartrees.}%
\begin{tabular}{c|ccc}
\hline\hline Molecule & E$_{\text{RHF}}$ & CCSD & DMRG(500)
\\ \hline
C$_{10}$H$_{8}$ & $-378.66597$ & $-378.85130$ & $-378.85360$\\
\hline C$_{14}$H$_{10}$ & $-529.44420$ & $-529.70634$ & $-529.71032$\\
\hline C$_{18}$H$_{12}$ & $-680.21823$ & $-680.56059$ & $-680.56538$\\
\hline C$_{22}$H$_{14}$ & $-830.99045$ & $-831.41614$ & $-831.42016$\\
\hline\hline
\end{tabular}%
\label{tab:Erhf}
\end{table*}

\begin{figure}[htp]
\includegraphics[scale=0.8]{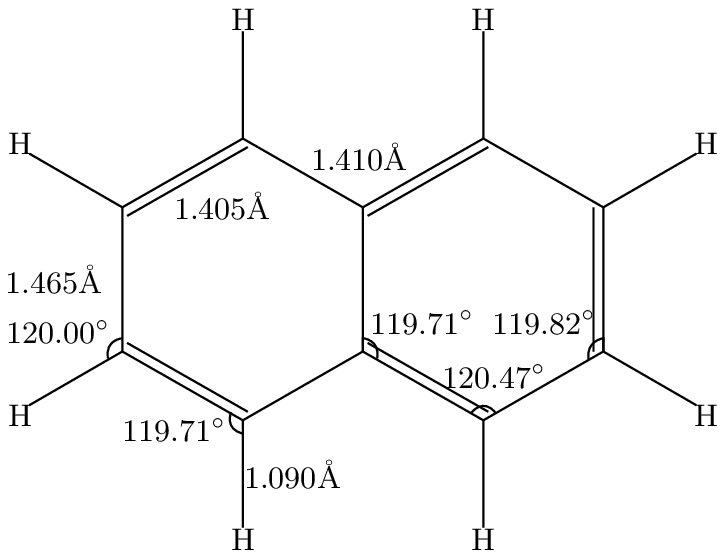}
\caption{Naphthalene model geometry.}\label{fig:geom}
\end{figure}

We used a model geometry for the acenes with $C_{2v}$  symmetry. The
C-H bond lengths were 1.090 {\AA}. Along the legs of the acene
ladder, the alternate C-C bond lengths were 1.410 {\AA} and 1.405
{\AA}, respectively. Along the rungs of the acene ladder, the C-C
bond length was 1.465 {\AA}. An example  geometry for naphthalene is
shown in Fig. \ref{fig:geom}.

All calculations used the Slater-Type-Orbitals fitted to 3 Gaussians
minimal basis set (STO-3G), consisting of 2s1p functions on C and 1s
functions on H \cite{hehre1969}. We obtained the atomic orbital
integrals and Restricted Hartree-Fock (RHF) orbitals from the
\textsc{psi3.2} package \cite{PSI3}. The RHF energies are given in
Table \ref{tab:Erhf}. For the excited state calculations, we used a
$\pi$-active space consisting of one $p_z$ orbital per carbon i.e.
$n$-acene would have a $(4n+2,4n+2)$ active space.  In the DMRG
calculations, we further symmetrically orthonormalized the $p_z$
orbitals with respect to the overlap $S$. This gave a local
orthonormal basis which yields faster convergence in the DMRG
calculations. The remaining non-active orbitals from the RHF
calculations were kept frozen in all calculations.

\begin{figure}[htp]
\includegraphics[scale=.8]{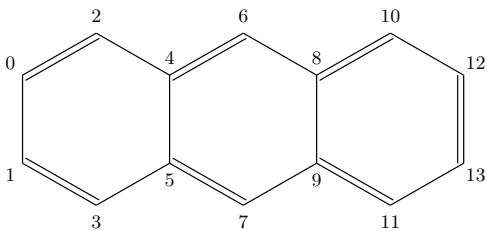}
\caption{The orbital ordering used for anthracene.}\label{fig:order}
\end{figure}

We calculated excitation energies  with
the State-Averaged Davidson (SA-D),
Harmonic Davidson  (HD), and State-Averaged Harmonic Davidson
 (SA-HD) algorithms described in Sec. \ref{sec:the}.
Our calculations used the local quadratic-scaling DMRG algorithm
described in Ref. \cite{hachmann2006mcl}. We employed a screening
threshold of $10^{-8}$ Hartrees ($E_h$) with no spatial symmetry.
The ordering of the orbitals for anthracene is shown in Fig.
\ref{fig:order} and the other acenes were ordered similarly. In all
of our sweeps, we added a small amount of random noise ($10^{-6} -
10^{-8}$) to the density matrix so that we would not lose important
quantum numbers \cite{chan2004als, mitrushenkov2003qcu}. In the
current algorithm it is difficult to converge  DMRG energies beyond
the intrinsic accuracy associated with the finite number $M$ of DMRG
basis states. Thus DMRG energies were converged to within 1
milliHartree (m$E_h$) ($M=50$), 0.5 m$E_h$ ($M=100$), 0.5 m$E_h$
($M=250$), or 0.1 m$E_h$ ($M=500$), respectively. We note that our
largest $M$ DMRG excitation energies are essentially exact (within
the  one-particle basis) to all reported digits. This is possible
for the large active spaces used here because of the compact
parametrisation afforded by the DMRG wavefunction.

In the HD and SA-HD calculations, the shift $\omega$ for a specific root
was obtained as
follows. To begin, we guessed an initial shift (typically based on our previous SA calculations).
 In the case where the shift was too low or too high, the next guess for $\omega$ was
obtained  from the DMRG (block) iteration, where an undesired state
first appeared as the ground state of the Harmonic Davidson
procedure. The shift $\omega$ was then taken to lie on the correct
side of the desired state in this iteration. In this simple manner,
we found that we could obtain a suitable shift for a given root with
at most two to three guesses.

To determine the symmetries  of the
excitations in the DMRG calculations we used the following method. Firstly, spin
symmetries were obtained from the expectation value of $S^2$. To
obtain the spatial symmetries, we first assumed
that the ground-state $\Psi_0$ was of $A_1$ symmetry (as in
experiment). For the excited states, we examined "dipole" type matrix elements
$\langle \Psi_0|n_0^\alpha + n_0^\beta - n_1^\alpha - n_1^\beta|\Psi_i\rangle$  (essentially
a dipole transition element along the short-axis of the acene; 0 and 1
refer to atom labels in Fig. \ref{fig:order}.) For singlet excited states
a non-vanishing dipole then implied $B_2$ symmetry, while a vanishing
dipole implied $A_1$ symmetry. For the triplet
excited states, all such matrix elements vanish. However, we could
still determine the spatial symmetry through the expectation value
 $\langle \Psi_0|n_0^\alpha - n_1^\alpha|\Psi_i
\rangle$ since $n_0^\alpha - n_1^\alpha$ does not preserve
spin symmetry and creates a residual  expectation value
from which one can determine the spatial symmetry of the excited state.

To obtain the orbital character of the excitations, we
calculated transition one-particle density matrices $\langle
\Psi_0|a^\dag_i a_j|\Psi_i\rangle$, where $\Psi_i$ denotes the $i$th
excited state and identified the largest matrix elements.

We further calculated the excitation spectrum (in the same
$\pi$-active space as the DMRG calculations) with the
Equation-Of-Motion Coupled Cluster Singles and Doubles method
(EOM-CCSD) \cite{eomccsdreview} using the  \textsc{dalton}
package\cite{dalton}.

\subsection{Comparison of Excited-state algorithms for DMRG by SA, HD, and SA-HD}
\label{sub:SADMRG}

\begin{table*}[tbp]
\caption{DMRG excitation energies for naphthalene (C$_{10}$H$_{8}$)
obtained with the SA-D, HD and SA-HD
  algorithms. All energies are in eV. State 0 refers to the
  ground-state, and SA[m-n] refers to a state-average over all states
  from the $m$th to $n$th excited state. Numbers in parentheses give
  the number of DMRG states $M$. The ``Exact (HD(500))'' numbers are the
  (near-exact) excitation energies, while  other entries give the
  errors from  this  result. The ``Excitation'' row gives the
  character of the excitation where 1 denotes HOMO, 2 denotes HOMO-1,
  $1^\prime$ denotes LUMO, $2^\prime$ denotes LUMO+1 and so on. The last column gives the mean
  improvement in the excitation energy over the SA [0-7] D result with the
  same $M$. n.c. denotes no convergence. }%
\begin{tabular}{c||c|c|c|c|c|c|c|c|c}
\hline\hline Method & \multicolumn{8}{c}{State}& Mean  \\
& $1^1A_1$ & $1^3B_2$ & $2^1A_1$ & $1^3A_1$ & $2^3B_2$ & $3^3B_2$ &
$2^3A_1$ & $3^1A_1$
 & Improvement
\\ \hline
\multirow{2}{*}{Excitation} &$      $&$ 1\rightarrow 1^\prime   $&$ 2\rightarrow 1^\prime   $&$ 2\rightarrow 1^\prime   $&$ 2\rightarrow 2^\prime   $&$ 3\rightarrow 1^\prime   $&$ 4\rightarrow 1^\prime   $&$ 3\rightarrow 2^\prime   $&$     $ \\
    &$      $&$ 2\rightarrow 2^\prime   $&$ 1\rightarrow 2^\prime   $&$ 1\rightarrow 2^\prime   $&$ 1\rightarrow 1^\prime   $&$ 1\rightarrow 3^\prime   $&$ 1\rightarrow 4^\prime   $&$ 2\rightarrow 3^\prime   $&$     $ \\    \hline
Exact (HD(500))  &$  0.00   $&$ 2.86    $&$ 4.08    $&$ 4.34
$&$ 4.63    $&$ 4.70    $&$ 5.51    $&$ 5.87    $&$     $ \\
\hline
SA [0-7] D (50)   &$  0.13    $&$ 0.09    $&$ 0.21    $&$ 0.46    $&$ 0.18    $&$ 0.15    $&$ 0.26    $&$  0.19   $&$     $ \\
SA [0-3] D (50)    &$  0.11    $&$ 0.17    $&$ 0.18    $&$ 0.34    $&$     $&$     $&$     $&$     $&$ 0.02    $ \\
SA [3-7] HD (50)    &$      $&$     $&$     $&$ 0.46    $&$ 0.17    $&$ 0.18    $&$ 0.24    $&$ 0.25    $&$ -0.01    $ \\
HD (50) &$  0.04    $&$ 0.05    $&$ 0.08    $&$ \text{n.c.}$&$
\text{n.c.}  $&$ \text{n.c.}  $&$ \text{n.c.} $&$ 0.08    $&$ 0.09 $ \\
SA [2-3] HD (50)    &$      $&$     $&$ 0.22    $&$ 0.25    $&$     $&$     $&$     $&$     $&$ 0.10    $ \\
\hline
SA [0-7] D (100)  &$  0.01    $&$ 0.02    $&$ 0.02    $&$ 0.03    $&$ 0.02    $&$ 0.02    $&$ 0.02    $&$ 0.02    $&$  $ \\
SA [0-3] D (100)   &$  0.01    $&$ 0.01    $&$ 0.01    $&$ 0.02    $&$     $&$     $&$     $&$     $&$ 0.01    $ \\
SA [3-7] HD (100)   &$      $&$     $&$     $&$ 0.03    $&$ 0.02    $&$ 0.02    $&$ 0.02    $&$ 0.02    $&$ 0.00    $ \\
HD (100)    &$  0.00    $&$ 0.01    $&$ 0.01    $&$ 0.01    $&$ 0.01
$&$ \text{n.c}  $&$ 0.01    $&$ 0.01    $&$ 0.01    $ \\    \hline
SA [0-7] D (250)  &$  0.00    $&$ 0.00    $&$ 0.00    $&$ 0.00    $&$ 0.00    $&$ 0.00    $&$ 0.00    $&$ 0.02    $&$  $ \\
SA [0-3] D (250)   &$  0.00    $&$ 0.00    $&$ 0.00    $&$ 0.00    $&$     $&$     $&$     $&$     $&$ 0.00    $ \\
SA [3-7] HD (250)   &$      $&$     $&$     $&$ 0.00    $&$ 0.00    $&$ 0.00    $&$ 0.00    $&$ 0.00    $&$ 0.00    $ \\
HD (250)    &$  0.00    $&$ 0.00    $&$ 0.00    $&$ 0.00    $&$ 0.00
$&$ 0.00    $&$ 0.00    $&$ 0.00    $&$ 0.00    $ \\    \hline
SA [0-7] D (500)  &$  0.00    $&$ 0.00    $&$ 0.00    $&$ 0.00    $&$ 0.00    $&$ 0.00    $&$ 0.00    $&$ 0.00    $&$    $ \\
SA [0-3] D (500)   &$  0.00    $&$ 0.00    $&$ 0.00    $&$ 0.00    $&$     $&$     $&$     $&$     $&$ 0.00    $ \\
SA [3-7] HD (500)   &$      $&$     $&$     $&$ 0.00    $&$ 0.00    $&$ 0.00    $&$ 0.00    $&$ 0.00    $&$ 0.00    $ \\
 \hline\hline
\end{tabular}%
\label{tab:fin}
\end{table*}

\begin{table*}[tbp]
\caption{DMRG excitation energies for anthracene
(C$_{14}$H$_{10}$). Refer to table \ref{tab:fin} for details.}
\begin{tabular}{c||c|c|c|c|c|c|c|c|c}
\hline\hline Method & \multicolumn{8}{c}{State}& Mean \\
& $1^1A_1$ & $1^3B_2$ & $2^1A_1$ & $2^3B_2$ & $1^3A_1$ & $3^3B_2$ &
$2^3A_1$ & $3^1A_1$
 & Improvement
\\ \hline
\multirow{3}{*}{Excitation} &$      $&$ 1\rightarrow 1^\prime   $&$ 2\rightarrow 1^\prime   $&$ 3\rightarrow 1^\prime   $&$ 2\rightarrow 1^\prime   $&$ 2\rightarrow 2^\prime   $&$ 4\rightarrow 1^\prime   $&$ 2\rightarrow 3^\prime   $&$     $ \\
    &$      $&$ 2\rightarrow 2^\prime   $&$ 1\rightarrow 2^\prime   $&$ 1\rightarrow 3^\prime   $&$ 1\rightarrow 2^\prime   $&$     $&$ 1\rightarrow 4^\prime   $&$ 3\rightarrow 2^\prime   $&$     $ \\
    &$      $&$ 3\rightarrow 3^\prime   $&$     $&$     $&$     $&$     $&$     $&$     $&$     $ \\    \hline
Exact(HD(500))  &$  0.00   $&$ 2.08    $&$ 3.57    $&$ 3.71
$&$ 3.85    $&$ 4.46    $&$ 4.73    $&$ 4.80    $&$     $ \\
\hline
SA [0-7] D (50)   &$  0.40    $&$ 0.45    $&$ 0.75    $&$ 0.65    $&$ 1.28    $&$ 0.77    $&$ 0.82    $&$ 0.91    $&$     $ \\
SA [0-3] D (50)    &$  0.29    $&$ 0.24    $&$ 0.46    $&$ 0.41    $&$     $&$     $&$     $&$     $&$ 0.21    $ \\
SA [3-7] HD (50)    &$      $&$     $&$     $&$ 0.73    $&$ 0.58    $&$ 0.60    $&$ 0.47    $&$ 0.69    $&$ 0.27    $ \\
HD (50) &$  0.12    $&$ 0.13    $&$ 0.40    $&$ \text{n.c.} $&$
\text{n.c.} $&$ \text{n.c.} $&$ \text{n.c.} $&$ \text{n.c.} $&$ 0.32
$ \\
SA [2-3] HD (50)    &$      $&$     $&$ 0.49    $&$ 0.41    $&$     $&$     $&$     $&$     $&$ 0.25    $ \\
   \hline
SA [0-7] D (100)  &$  0.12    $&$ 0.12    $&$ 0.15    $&$ 0.15    $&$ 0.25    $&$ 0.23    $&$ 0.18    $&$ 0.19    $&$     $ \\
SA [0-3] D (100)   &$  0.07    $&$ 0.07    $&$ 0.10    $&$ 0.09    $&$     $&$     $&$     $&$     $&$ 0.05    $ \\
SA [3-7] HD (100)   &$      $&$     $&$     $&$ 0.15    $&$ 0.20    $&$ 0.20    $&$ 0.16    $&$ 0.28    $&$ 0.00    $ \\
HD (100)    &$  0.01    $&$ 0.03    $&$ 0.05    $&$ 0.04    $&$
\text{n.c.} $&$\text{n.c.} $&$\text{n.c.} $&$\text{n.c.} $&$ 0.10    $ \\
SA [5-6] HD (100)   &$      $&$     $&$     $&$     $&$     $&$ 0.13    $&$ 0.22    $&$     $&$ 0.03    $ \\
SA [6-7] HD (100)   &$      $&$     $&$     $&$     $&$     $&$     $&$ 0.14    $&$ 0.12    $&$ 0.09    $ \\
\hline
SA [0-7] D (250)  &$  0.01    $&$ 0.01    $&$ 0.01    $&$ 0.02    $&$ 0.03    $&$ 0.02    $&$ 0.02    $&$ 0.02    $&$     $ \\
SA [0-3] D (250)   &$  0.00    $&$ 0.00    $&$ 0.01    $&$ 0.01    $&$     $&$     $&$     $&$     $&$ 0.01    $ \\
SA [3-7] HD (250)   &$      $&$     $&$     $&$ 0.02    $&$ 0.03    $&$ 0.02    $&$ 0.02    $&$ 0.09    $&$ -0.01    $ \\
HD (250)    &$  0.00    $&$ 0.00    $&$ 0.00    $&$ 0.00    $&$ 0.00
$&$ 0.00    $&$ 0.00    $&$ 0.00    $&$ 0.02    $ \\    \hline
SA [0-7] D (500)  &$  0.00    $&$ 0.00    $&$ 0.00    $&$ 0.00    $&$ 0.00    $&$ 0.00    $&$ 0.00    $&$ 0.00    $&$    $ \\
SA [0-3] D (500)   &$  0.00    $&$ 0.00    $&$ 0.00    $&$ 0.00    $&$     $&$     $&$     $&$     $&$ 0.00    $ \\
SA [3-7] HD (500)   &$      $&$     $&$     $&$ 0.00    $&$ 0.00    $&$ 0.00    $&$ 0.00    $&$ 0.00    $&$ 0.00    $ \\
 \hline \hline
\end{tabular}%
\label{tab:fin2}
\end{table*}

\begin{table*}[tbp]
\caption{DMRG excitation energies  for naphthacene
(C$_{18}$H$_{12}$). Refer to table \ref{tab:fin} for details.}%
\begin{tabular}{c||c|c|c|c|c|c|c|c|c}
\hline\hline Method & \multicolumn{8}{c}{State}& Mean \\
& $1^1A_1$ & $1^3B_2$ & $2^3B_2$ & $2^1A_1$ & $1^3A_1$ & $3^1A_1$ &
$3^3B_2$ & $2^3A_1$ & Improvement
\\ \hline
\multirow{3}{*}{Excitation} &$      $&$ 1\rightarrow 1^\prime   $&$ 3\rightarrow 1^\prime   $&$ 2\rightarrow 1^\prime   $&$ 2\rightarrow 1^\prime   $&$ 3\rightarrow 2^\prime   $&$ 5\rightarrow 1^\prime   $&$ 4\rightarrow 1^\prime   $&$     $ \\
    &$      $&$ 3\rightarrow 3^\prime   $&$ 1\rightarrow 3^\prime   $&$ 1\rightarrow 2^\prime   $&$ 1\rightarrow 2^\prime   $&$ 2\rightarrow 3^\prime   $&$ 1\rightarrow 5^\prime   $&$ 1\rightarrow 4^\prime   $&$     $ \\    \hline
Exact(HD(500))  &$  0.00   $&$ 1.52    $&$ 2.95    $&$ 3.27
$&$ 3.50    $&$ 3.93    $&$ 4.02    $&$ 4.23    $&$     $ \\
\hline
SA [0-7] D (50)   &$  0.71    $&$ 0.81    $&$ 1.07    $&$ 1.33    $&$ 1.75    $&$ 1.45    $&$ 1.51    $&$ 1.70    $&$     $ \\
SA [0-3] D (50)    &$  0.50    $&$ 0.48    $&$ 0.66    $&$ 0.92    $&$     $&$     $&$     $&$     $&$ 0.34    $ \\
SA [2-7] HD (50)    &$      $&$     $&$ 0.20    $&$ 1.04    $&$ 1.25    $&$ 1.39    $&$ 1.34    $&$ 1.34    $&$ 0.38    $ \\
SA [3-7] HD (50)    &$      $&$     $&$     $&$ \text{n.c.} $&$ \text{n.c.} $&$ \text{n.c.} $&$ \text{n.c.} $&$ \text{n.c.} $&$ 0.00    $ \\
HD (50) &$  0.20    $&$ 0.27    $&$ \text{n.c.} $&$ \text{n.c.} $&$
\text{n.c.} $&$ \text{n.c.} $&$
\text{n.c.} $&$ \text{n.c.} $&$ 0.52 $ \\
SA [1-2] HD (50)    &$      $&$ 1.43    $&$ 0.95    $&$     $&$     $&$     $&$     $&$     $&$ -0.25    $ \\
SA [2-3] HD (50)    &$      $&$     $&$ 0.67    $&$ 0.84    $&$     $&$     $&$     $&$     $&$ 0.45    $ \\
\hline
SA [0-7] D (100)  &$  0.27    $&$ 0.28    $&$ 0.32    $&$ 0.40    $&$ 0.58    $&$ 0.47    $&$ 0.41    $&$ 0.50    $&$     $ \\
SA [0-3] D (100)   &$  0.12    $&$ 0.14    $&$ 0.16    $&$ 0.20    $&$     $&$     $&$     $&$     $&$ 0.16    $ \\
SA [2-7] HD (100)   &$      $&$     $&$ 0.33    $&$ 0.40    $&$ 0.57    $&$ 0.45    $&$ 0.39    $&$ 0.54    $&$ 0.00    $ \\
SA [3-7] HD (100)   &$      $&$     $&$     $&$ \text{n.c.} $&$ \text{n.c.} $&$ \text{n.c.} $&$ \text{n.c.} $&$ \text{n.c.} $&$ 0.00    $ \\
HD (100)    &$  0.03    $&$ 0.06    $&$ 0.08    $&$ 0.11    $&$
\text{n.c.} $&$ \text{n.c.} $&$ \text{n.c.} $&$ \text{n.c.} $&$ 0.25
$
\\    \hline
SA [0-7] D (250)  &$  0.03    $&$ 0.04    $&$ 0.04    $&$ 0.06    $&$ 0.08    $&$ 0.08    $&$ 0.06    $&$ 0.07    $&$    $ \\
SA [0-3] D (250)   &$  0.01    $&$ 0.01    $&$ 0.02    $&$ 0.03    $&$     $&$     $&$     $&$     $&$ 0.03    $ \\
SA [3-7] HD (250)   &$      $&$     $&$     $&$ 0.05    $&$ 0.07    $&$ 0.07    $&$ 0.06    $&$ 0.09    $&$ 0.00    $ \\
HD (250)    &$  0.00    $&$ 0.00    $&$ 0.00    $&$ \text{n.c.} $&$
\text{n.c.} $&$ 0.02    $&$ 0.01    $&$ 0.01    $&$ 0.05    $ \\
\hline
SA [0-7] D (500)  &$  0.00    $&$ 0.00    $&$ 0.00    $&$ 0.01    $&$ 0.01    $&$ 0.01    $&$ 0.01    $&$ 0.01    $&$    $ \\
SA [0-3] D (500)   &$  0.00    $&$ 0.00    $&$ 0.00    $&$ 0.00    $&$     $&$     $&$     $&$     $&$ 0.00    $ \\
SA [3-7] HD (500)   &$      $&$     $&$     $&$ 0.01    $&$ 0.01    $&$ 0.01    $&$ 0.01    $&$ 0.01    $&$ 0.00    $ \\
\hline \hline
\end{tabular}%
\label{tab:fin3}
\end{table*}

\begin{table*}[tbp]
\caption{DMRG excitation energies  for pentacene
(C$_{22}$H$_{14}$). Refer to table \ref{tab:fin} for details.}%
\begin{tabular}{c||c|c|c|c|c|c|c|c|c}
\hline\hline Method & \multicolumn{8}{c}{State}& Mean \\
& $1^1A_1$ & $1^3B_2$ & $2^3B_2$ & $2^1A_1$ & $1^3A_1$ & $3^1A_1$ &
$3^3B_2$ & $2^3A_1$ & Improvement
\\ \hline
\multirow{3}{*}{Excitation} &$      $&$ 1\rightarrow 1^\prime   $&$ 2\rightarrow 1^\prime   $&$ 3\rightarrow 1^\prime   $&$ 3\rightarrow 1^\prime   $&$ 3\rightarrow 2^\prime   $&$ 4\rightarrow 1^\prime   $&$ 1\rightarrow 5^\prime   $&$     $ \\
    &$      $&$ 2\rightarrow 2^\prime   $&$ 1\rightarrow 2^\prime   $&$ 1\rightarrow 3^\prime   $&$ 1\rightarrow 3^\prime   $&$ 2\rightarrow 3^\prime   $&$ 1\rightarrow 4^\prime   $&$ 5\rightarrow 1^\prime   $&$     $ \\
    &$      $&$     $&$     $&$     $&$     $&$     $&$ 2\rightarrow 2^\prime   $&$     $&$     $ \\    \hline
Exact(HD(500))  &$  0.00   $&$ 1.15    $&$ 2.39    $&$ 3.10
$&$ 3.15    $&$ 3.30    $&$ 3.43    $&$ 3.88    $&$     $ \\
\hline
SA [0-7] D (50)   &$  1.10    $&$ 1.55    $&$ 1.79    $&$ 1.86    $&$ 2.31    $&$ 2.29    $&$ 2.26    $&$ 2.38    $&$     $ \\
SA [0-3] D (50)    &$  0.72    $&$ 0.72    $&$ 0.98    $&$ 1.24    $&$     $&$     $&$     $&$     $&$ 0.66    $ \\
SA [2-7] HD (50)    &$      $&$     $&$ 1.23    $&$ 1.62    $&$ 1.88    $&$ 1.75    $&$ 1.97    $&$ 1.95    $&$ 0.41    $ \\
SA [3-7] HD (50)    &$      $&$     $&$     $&$ \text{n.c.} $&$ \text{n.c.} $&$ \text{n.c.} $&$ \text{n.c.} $&$ \text{n.c.} $&$ 0.00    $ \\
HD (50) &$  0.29    $&$ 0.40    $&$ \text{n.c.} $&$ \text{n.c.} $&$
\text{n.c.} $&$
\text{n.c.} $&$ \text{n.c.} $&$ \text{n.c.} $&$ 0.98 $ \\
SA [2-3] HD (50)    &$      $&$     $&$ 0.68    $&$ 1.21    $&$     $&$     $&$     $&$     $&$ 0.88    $ \\
\hline
SA [0-7] D (100)  &$  0.44    $&$ 0.48    $&$ 0.52    $&$ 0.70    $&$ 0.87    $&$ 0.80    $&$ 0.82    $&$ 0.87    $&$     $ \\
SA [0-3] D (100)   &$  0.31    $&$ 0.33    $&$ 0.34    $&$ 0.38    $&$     $&$     $&$     $&$     $&$ 0.20    $ \\
SA [2-7] HD (100)   &$      $&$     $&$ 0.47    $&$ 0.56    $&$ 0.75    $&$ 0.67    $&$ 0.56    $&$ 0.69    $&$ 0.15    $ \\
SA [3-7] HD (100)   &$      $&$     $&$     $&$ \text{n.c.} $&$ \text{n.c.} $&$ \text{n.c.} $&$ \text{n.c.} $&$ \text{n.c.} $&$ 0.00    $ \\
HD (100)    &$  0.04    $&$ 0.09    $&$ 0.14    $&$ \text{n.c.} $&$
\text{n.c.} $&$ \text{n.c.} $&$ \text{n.c.} $&$ \text{n.c.} $&$ 0.39
$ \\    \hline
SA [0-7] D (250)  &$  0.06    $&$ 0.08    $&$ 0.10    $&$ 0.12    $&$ 0.15    $&$ 0.16    $&$ 0.12    $&$ 0.15    $&$     $ \\
SA [0-3] D (250)   &$  0.02    $&$ 0.02    $&$ 0.03    $&$ 0.04    $&$     $&$     $&$     $&$     $&$ 0.06    $ \\
SA [3-7] HD (250)   &$      $&$     $&$     $&$ 0.09    $&$ 0.10    $&$ 0.12    $&$ 0.10    $&$ 0.12    $&$ 0.03    $ \\
HD (250)    &$  0.00    $&$ 0.01    $&$ 0.01    $&$ 0.02    $&$
\text{n.c.} $&$ \text{n.c.} $&$ \text{n.c.} $&$ \text{n.c.} $&$ 0.08
$ \\    \hline
SA [0-7] D (500)  &$  0.00    $&$ 0.01    $&$ 0.01    $&$ 0.01    $&$ 0.02    $&$ 0.02    $&$ 0.02    $&$ 0.02    $&$    $ \\
SA [0-3] D (500)   &$  0.00    $&$ 0.00    $&$ 0.00    $&$ 0.00    $&$     $&$     $&$     $&$     $&$ 0.01    $ \\
SA [3-7] HD (500)   &$      $&$     $&$     $&$ 0.01    $&$ 0.01    $&$ 0.02    $&$ 0.02    $&$ 0.02    $&$ 0.00    $ \\
\hline \hline
\end{tabular}%
\label{tab:fin4}
\end{table*}

\begin{table*}[tbp]
\caption{DMRG excitation energies  for the higher excited states
of naphthalene (C$_{10}$H$_{8}$). Refer to table \ref{tab:fin} for details.}%
\begin{tabular}{c||c|c|c|c|c|c|c}
\hline\hline Method & \multicolumn{6}{c}{State} & Mean \\
& $2^3A_1$ & $3^1A_1$ & $4^3B_2$ & $3^3A_1$ & $1^1B_2$ & $4^1A_1$ &
Improvement
\\ \hline
\multirow{4}{*}{Excitation} &$  4\rightarrow 1^\prime   $&$ 3\rightarrow 2^\prime   $&$ 4\rightarrow 2^\prime   $&$ 2\rightarrow 3^\prime   $&$ 1\rightarrow 3^\prime   $&$ 4\rightarrow 1^\prime   $&$     $ \\
    &$  1\rightarrow 4^\prime   $&$ 2\rightarrow 3^\prime   $&$ 2\rightarrow 4^\prime   $&$ 3\rightarrow 2^\prime   $&$ 3\rightarrow 1^\prime   $&$ 1\rightarrow 4^\prime   $&$     $ \\
    &$      $&$     $&$     $&$     $&$ 4\rightarrow 2^\prime   $&$     $&$     $ \\
    &$      $&$     $&$     $&$     $&$ 2\rightarrow 4^\prime   $&$     $&$     $ \\    \hline
Exact(HD(500))  &$  5.51    $&$ 5.87    $&$ 6.28    $&$ 6.48    $&$
6.84    $&$ 6.84    $&$     $ \\    \hline
SA [0-11] D (50)  &$  0.29    $&$ 0.21    $&$ 0.19    $&$ 0.45    $&$ 0.32    $&$ 0.43    $&$     $ \\
SA [6-11] HDa (50)   &$  0.29    $&$ 0.20    $&$ 0.17    $&$ 0.46    $&$ 0.35    $&$ 0.43    $&$ 0.00    $ \\
HD (50) &$  \text{n.c}  $&$ 0.08    $&$ \text{n.c}  $&$ \text{n.c}
$&$ \text{n.c}  $&$ \text{n.c}  $&$ 0.13    $ \\
\hline
SA [0-11] D (100) &$  0.03    $&$ 0.03    $&$ 0.03    $&$ 0.03    $&$ 0.05    $&$ 0.06    $&$     $ \\
SA [6-11] HDa (100)  &$  0.03    $&$ 0.02    $&$ 0.03    $&$ 0.03    $&$ 0.05    $&$ 0.06    $&$ 0.00    $ \\
HD (100)    &$  0.01    $&$ 0.01    $&$ 0.01    $&$ 0.01    $&$ 0.02
$&$ \text{n.c}  $&$ 0.02    $ \\
SA [10-11] HDa (100) &$      $&$     $&$     $&$     $&$ 0.04    $&$
0.04    $&$ 0.02    $ \\
 \hline
SA [0-11] D (250) &$  0.00    $&$ 0.00    $&$ 0.00    $&$ 0.00    $&$ 0.00    $&$ 0.00    $&$     $ \\
SA [6-11] HDa (250)  &$  0.00    $&$ 0.00    $&$ 0.00    $&$ 0.00    $&$ 0.00    $&$ 0.00    $&$ 0.00    $ \\
HD (250)    &$  0.00    $&$ 0.00    $&$ 0.00    $&$ 0.00    $&$ 0.00
$&$ \text{n.c}  $&$ 0.00    $ \\
SA [10-11] HDa (250) &$      $&$     $&$     $&$     $&$ 0.00    $&$ 0.00    $&$ 0.00    $ \\
 \hline
SA [0-11] D (500) &$  0.00    $&$ 0.00    $&$ 0.00    $&$ 0.00    $&$ 0.00    $&$ 0.00    $&$     $ \\
SA [6-11] HDa (500)  &$  0.00    $&$ 0.00    $&$ 0.00    $&$ 0.00    $&$ 0.00    $&$ 0.00    $&$ 0.00    $ \\
\hline \hline
\end{tabular}
\label{tab:fin5}
\end{table*}

The ground state DMRG energies for the acenes are given in Table
\ref{tab:Erhf}. Tables \ref{tab:fin}, \ref{tab:fin2},
\ref{tab:fin3}, and \ref{tab:fin4} contain the first seven $\pi -
\pi^*$ excitation energies for each acene, while Fig. \ref{fig:ccsd}
displays them in graphical form. Under $C_{2v}$ symmetry, the only
two possible representations  of the $\pi - \pi^*$ excited states
are $^{1,3}A_1$ and $^{1,3}B_2$. Experimentally, there are three
well-documented singlet bands that appear in the visible spectrum
\cite{kadantsev2006,heinze2000}. The $\alpha$-band and $\beta$-band
correspond to a polarization along the long axis and the $p$-band
corresponds to a transverse polarization. We observed the
$\alpha$-transition as the lowest singlet excitation in each acene.
Neither the $p$-band nor the $\beta$-band appeared within the first
eight states of each acene. Instead, for the case of naphthalene,
the $p$-band emerged at 8.42 eV (state 19). The $p$-band normally
appears lower in the spectrum, but the absence of  dynamic
$\sigma-\pi$ correlations is responsible for its artificially high
excitation energy here. This is consistent with previous studies of
acenes using Complete-Active-Space Self-Consistent-Field (CASSCF)
and Complete-Active-Space Moller-Plesset second order perturbation
theory (CASMP2) theory \cite{nakatsuji1987,
kawashima1999,hashimoto1996}. Triplet excitations are somewhat
harder to measure experimentally. We observe that the triplet
excitation energies decrease in energy more rapidly with system size
than the singlet excitations. Thus while in naphthalene and
anthracene there is one triplet level between the first two singlet
excitations, in naphthacene and pentacene there are two.

Comparing the accuracies of the SA-D, HD, and SA-HD calculations we
observe that as expected, (other than by the size of the DMRG basis
$M$), the accuracy in the excitation energies is determined
primarily by the number of eigenvectors in the state-average.
Consequently the traditional SA-D algorithm  yielded  the lowest
accuracy (as it averages over all states between the ground state
and excited state of interest) while the HD calculations were
correspondingly the most accurate since they targeted a single state
at a time. The accuracy of the SA-HD calculations lay somewhere in
between depending on the number of states used in the average. In
all cases, the differences between the various algorithms was most
marked for the smaller sizes $M$ of the DMRG basis, as for larger
$M$ all the wavefunctions become essentially exact. We would expect
the differences to become more pronounced in larger systems, where
we are unable to use a sufficiently large $M$ to reach exactness.





Regarding the stabilities of the various algorithms, we found that
there were no difficulties  in converging the DMRG
sweeps to the correct states  with the SA-D algorithm. The HD
algorithm on the other hand exhibited the expected
convergence difficulties characteristic of
root-flipping for certain higher excited states.
As previously discussed, the stability of the HD algorithm would increase with
 the size of the DMRG many-body basis $M$.
In naphthacene, we required $M\geq250$ to converge
states 5-7 with the HD algorithm, while in pentacene, we required
$M=500$ to  converge states 4-7. While the HD algorithm exhibited  root-flipping,  it was ameliorated
with respect to simple eigenvector
following (defined as following the
$n^{\text{th}}$ eigenvector in the block Davidson algorithm in
successive DMRG iterations) because of the use of the shift  $\omega$. For example, with $M=100$, the third excited state of naphthalene  could not
be converged with simple eigenvector following, but could be
converged without difficulty using the HD algorithm.

Including a sufficient number of states in the SA-HD algorithm
 restored the stability of the convergence.
Certain ``competing'' states were particularly important
for the state average, especially for smaller $M$.
For all the acenes, the  second and third excited states were examples of such states. Thus while  the
 state averages SA[2-3] HD and SA[2-7] HD  converged without difficulty,
 calculations using SA[3-7] HD did not, at least for smaller $M$.


As mentioned previously, rather than choosing a shift to target specific excited states, we could take the different approach
of trying to find the excited states around the frequency of a given
shift $\omega$. In this way, we could piece together a complete
spectrum by performing, say, SA-HD or SA-HDa calculations with successively higher shifts.
To demonstrate this, we computed the excitation energies for
states 6-11 for naphthalene using the SA-HDa
algorithm with a shift chosen slightly above
the state 7 excitation energy as estimated from  the previous SA-HD [4-7]
calculation. These are shown in table \ref{tab:fin5}.

\subsection{Comparison of DMRG and EOM-CC excitation energies in the acenes}
\label{sub:CCDMRG}

\begin{figure}[p]
\begin{center}
\includegraphics[width=0.9\textwidth]{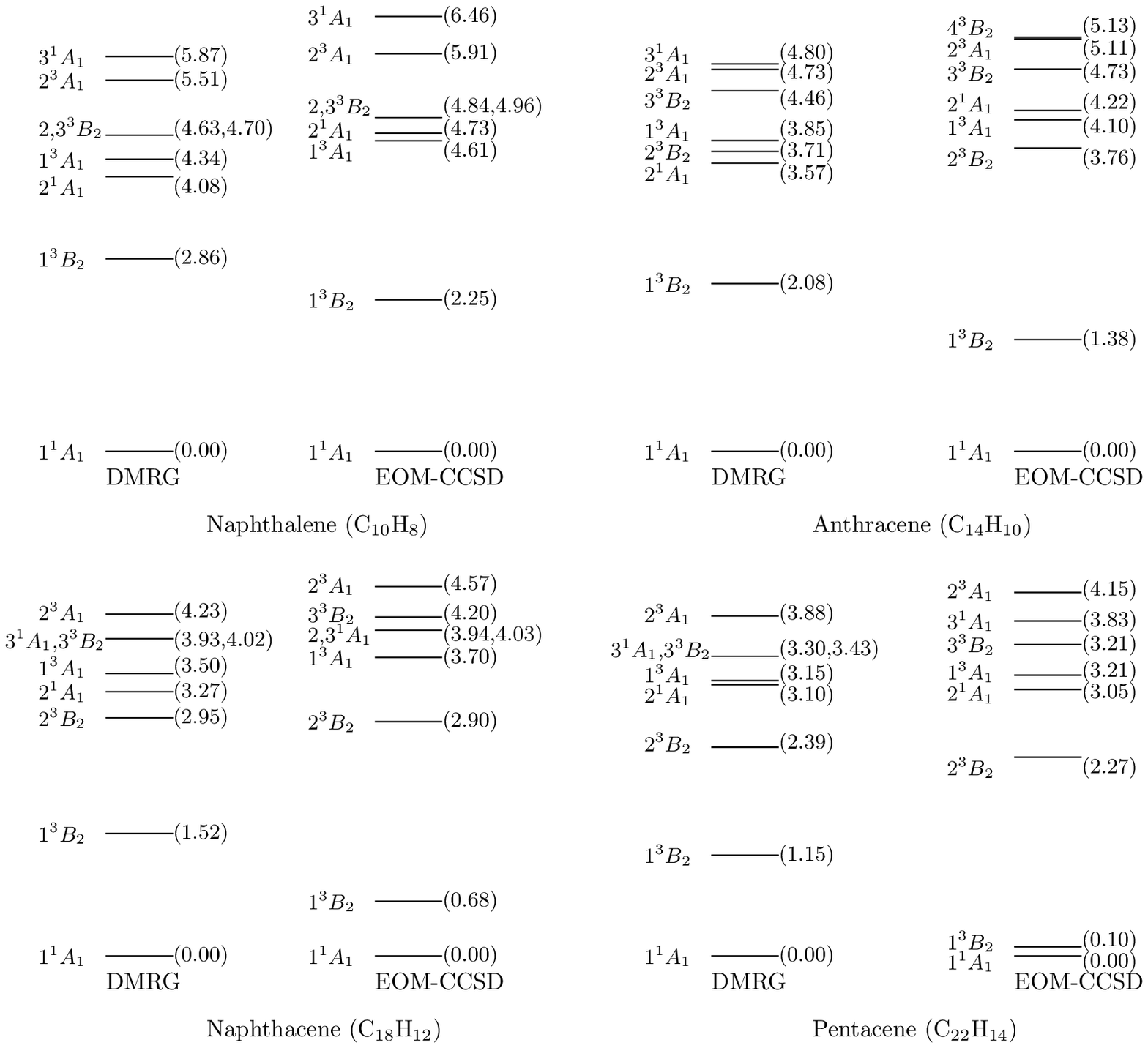}
\caption{Comparison of DMRG and EOM-CCSD excitation energies for
acenes.  All energies in eV.}\label{fig:ccsd}
\end{center}
\end{figure}

The ground state EOM-CCSD energies
for the acenes are summarized in Table \ref{tab:Erhf}.
We used our near-exact DMRG(500) excitation energies to examine the accuracy of
the EOM-CC method in acenes. The EOM-CCSD and the  DMRG
symmetries and excitation energies are shown in Fig.
\ref{fig:ccsd}. For the larger acenes, the EOM-CCSD excited states are in a
qualitatively different order as compared to DMRG. Similarly,
EOM-CCSD erroneously predicts a very small singlet-triplet
gap for the longer  acenes. This points to the necessity of including relatively high order
correlation effects to accurately describe excitations in the acenes.

\section{Conclusions}
\label{sec:con}

To overcome the computational and accuracy limitations of the traditional State-Averaged Davidson
Algorithm, which requires both  solving for and representing all states between the ground state and excited
state of interest, we have investigated a number of new
excited state algorithms within the context of
the Density Matrix Renormalization
Group (DMRG).  In the \textit{Harmonic Davidson} (HD) algorithm,  using  a shifted and
inverted operator enabled us to  directly solve for the
excited state of interest. In the \textit{State-Averaged
Harmonic Davidson} (SA-HD) algorithm, we combined the HD method with an average
over nearby excited states, to confer greater stability and overcome
problems of root-flipping in the non-linear optimisation of the wavefunction.

To assess the accuracy, stability, and computational cost of these
new methods we calculated  the low-lying excited states in the
acenes ranging from naphthalene to pentacene. We found that as
expected, in addition to the size of the DMRG basis $M$ used, the
accuracy was primarily determined by the number of states used in
the state average. Thus the State-Averaged Davidson approach gave
the least accuracy, the Harmonic Davidson algorithm, the highest,
and the State-Averaged Harmonic Davidson lay in between depending on
how many nearby states were included. The State-Averaged Harmonic
Davidson algorithm converged smoothly without root-flipping so long
as  nearby
 ``competing'' states were included in the average.

We also argued that through the shift $\omega$ in the Harmonic
Davidson algorithms we could piece together a complete excitation
spectrum by targeting different regions with successively higher
shifts. This we demonstrated by calculating some  higher lying
excited states in naphthalene.

Within the basis used, our DMRG excitation energies are near-exact
and we have used them  to assess the accuracy of the EOM-CCSD method
in the acenes.  We found that the EOM-CCSD excitation spectrum was
qualitatively  different from that of the DMRG for the larger acenes, which
demonstrates the necessity of including higher-order correlations to
properly describe the electronic spectrum of conjugated
quasi-one-dimensional molecules.

Finally, we observe that the Harmonic Davidson algorithms studied here
 are quite general methods and are not limited to the Density Matrix
 Renormalisation Group. Thus they may be useful also to target
  excited states in other multi-reference
theories, such as Complete Active Space Self-Consistent-Field
theory.

\begin{acknowledgments}
JH is funded by a Kekul\'{e} Fellowship of the Fond der Chemischen
Industrie (Fund of the German Chemical Industry). GKC acknowledges
support from Cornell University, Cornell Center for Materials
Research, the David and Lucile Packard Foundation in Science and
Engineering, and the National Science Foundation CAREER program
CHE-0645380.
\end{acknowledgments}

\bibliography{arxiv}

%
%
%
%
%

\end{document}